\documentclass[hyper,11pt,paper]{article}
\pdfoutput=1
\usepackage{jheppub}
\usepackage[usenames,dvipsnames,table]{xcolor}
\usepackage{graphicx,amsmath,amssymb,multirow,array,bm,mathrsfs}
\usepackage{epsf,amsfonts}
\usepackage[numbers,sort&compress]{natbib}
\title{
Black brane steady states
}
\author{Irene Amado and Amos Yarom}

\affiliation{Department of Physics, Technion, Haifa 32000, Israel}

\emailAdd{irene.r.amado@gmail.com}
\emailAdd{ayarom@physics.technion.ac.il}
\abstract{
We follow the evolution of an asymptotically AdS black brane with a fixed temperature gradient at spatial infinity until a steady state is formed. The resulting energy density and energy flux of the steady state in the boundary theory are compared to a conjecture on the behavior of steady states in conformal field theories. Very good agreement is found.  

}
\begin{document}
\maketitle

\section{Introduction }

Predicting the out of equilibrium behavior of many-body systems is a notoriously difficult problem. Many of the tools which have been developed to describe thermodynamic equilibrium are insufficient to explore systems which are far from it. Of the various non-equilibrium phenomenon, one that seems to have a tractable lead is the steady-state problem where one attempts to study the properties of a steady state which emerges once the system is placed between two heat baths. Recent progress on this problem has been seen in a variety of systems which may be described by a quantum field theory, see, e.g., \cite{Bernard:2012je,BD2013,Bernard:2013bqa,Joetalk,Chang:2013gba,Bhaseen:2013ypa,Freefermions,Bernard:2014qia}. In this work we focus on the steady-state generated when placing a $3$ space-time dimensional relativistic conformal gauge-theory between two heat baths.

When placing a system described by a two dimensional conformal field theory between two heat baths its late time, steady state, behavior is completely fixed in terms of the left and right central charges, the levels of the chiral currents (if present) and the relative pressure difference of the heat baths  \cite{Bernard:2012je,BD2013,Bernard:2013bqa,Joetalk,Chang:2013gba}. A recent conjecture which appeared in \cite{Joetalk,Chang:2013gba,Bhaseen:2013ypa} attempts to generalize this result to higher dimensional theories.

The form of the conjecture of \cite{Joetalk,Chang:2013gba,Bhaseen:2013ypa} is as follows. Let us consider two heat baths which are infinitely separated along one spatial direction, call it $z$, and are translation invariant along the remaining directions, $x_{\bot}$. Let us denote the pressure of the left and right heat baths by $P_L$ and $P_R$ respectively, and the relative pressure difference by
\begin{equation}
\label{E:deltap}
	\delta p = \frac{P_L-P_R}{P_L+P_R}\,.
\end{equation}
Without loss of generality we choose $P_L>P_R$.
Consider an initial state whose energy momentum tensor smoothly interpolates between the two heat baths and is also translation invariant along $x_{\bot}$. The mechanism by which energy is propagated in such a setting can be categorized into two types. A flow driven (or ballistic) mechanism by which the energy propagates via sound modes or their non-linear counterpart, and diffusion. If the system is flow driven then we may expect that at late times the energy of the system will propagate with a velocity $v_L$ towards the left heat bath and  with velocity $v_R$ towards the right heat bath generating a steady-state in between. If, in addition, the system thermalizes sufficiently rapidly so that the disturbance immediately behind the propagating waves is in thermal equilibrium, then energy-momentum conservation fixes the expectation value of the pressure, $T^{zz}$, and the energy flux along the $z$ direction, $T^{tz}$, at late times \cite{Joetalk,Chang:2013gba,Bhaseen:2013ypa}.

More precisely, in \cite{Chang:2013gba} it was shown that the aforementioned physical considerations about the evolution of the system leads to two branches of solutions for the possible late time behavior of $T^{zz}$ and $T^{tz}$. In \cite{Joetalk,Bhaseen:2013ypa} it was argued that a translation invariant steady state allows for only one possible late time solution. The work of \cite{Chang:2013gba} and \cite{Joetalk,Bhaseen:2013ypa} are consistent. 

\begin{figure}[hbt]
\centering{
\includegraphics[width=0.495\textwidth]{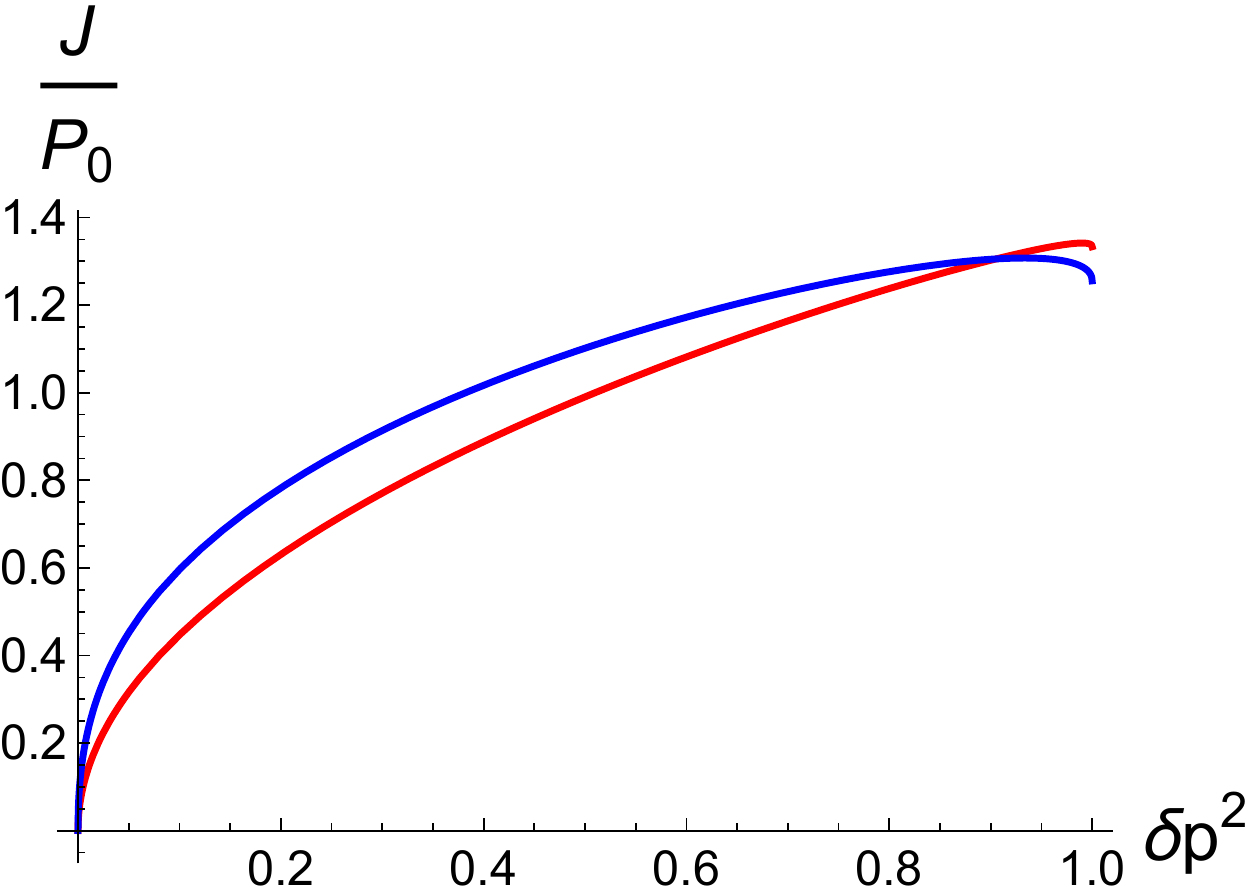}\hfill
\includegraphics[width=0.495\textwidth]{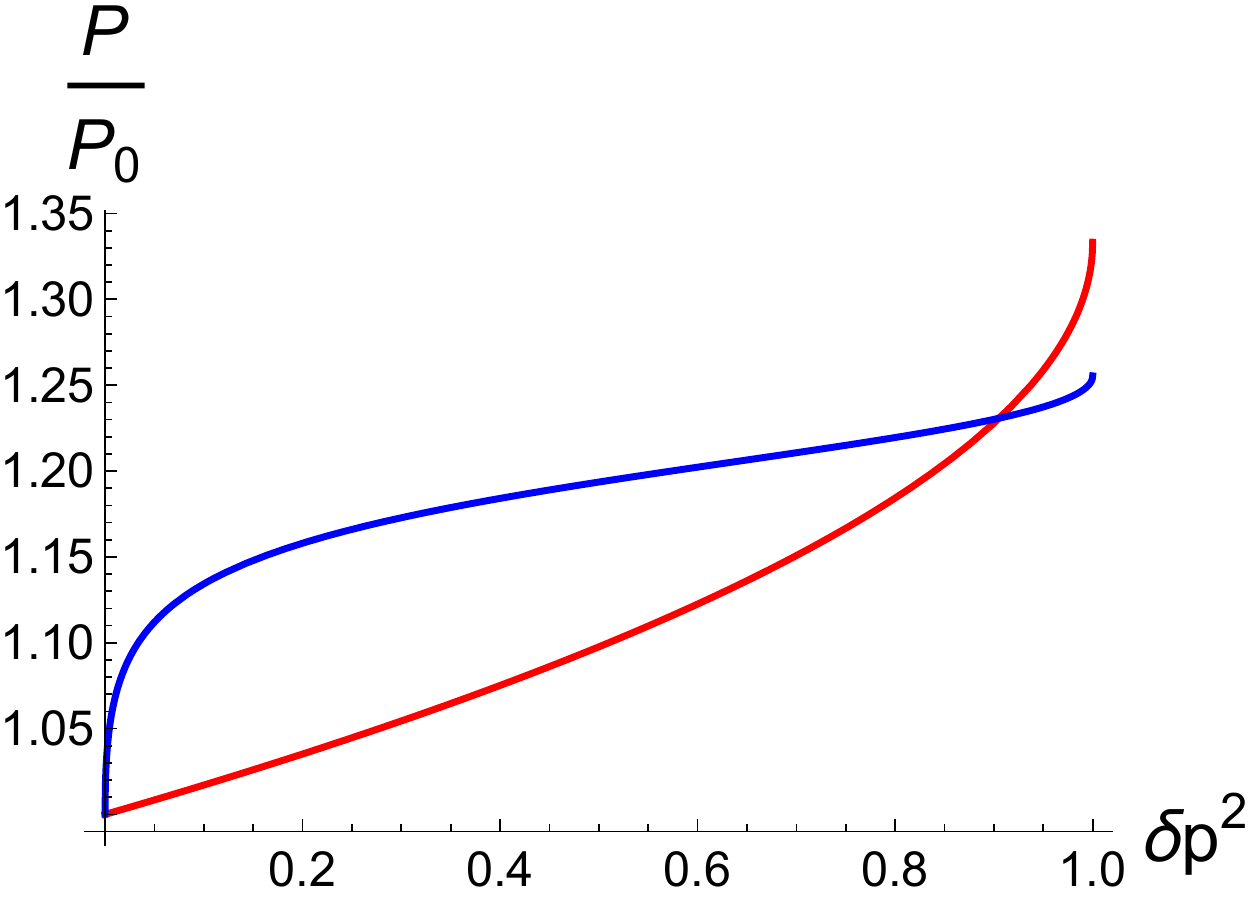}
\caption{ \label{F:3dsol} A plot of the conjectured universal values of the late time steady state energy flux (left panel) and pressure (right panel). Here $P_0$ is the average pressure of the heat baths, given by \eqref{E:P0def}, and $\delta p$ is the relative pressure difference, given by \eqref{E:deltap}. The red curves correspond to the thermodynamic branch of steady-states where the expectation value of the energy momentum tensor is equivalent to that of a boosted thermally equilibrated configuration. The blue curve corresponds to the ``other'' branch of steady-state configurations whose physical role is somewhat obscure.}
}
\end{figure}

In 3 space-time dimensions, the translation invariant steady-state (referred to as the thermodynamic branch in \cite{Chang:2013gba}) is characterized by
\begin{align}
\begin{split}
\label{E:redbranch}
	\frac{P}{P_0} \equiv \lim_{t\to\infty}\frac{\hbox{Tr}\left(\rho T^{zz} \right)}{P_0} &= \frac{1}{3} \left(4-\sqrt{1-\delta p^2}\right) \\
	\frac{J}{P_0}  \equiv \lim_{t\to\infty}\frac{\hbox{Tr}\left(\rho T^{tz} \right)}{P_0} & =  \frac{2}{3} \sqrt{5 \delta p^2 + \sqrt{1-\delta p^2} - 1}
\end{split}
\end{align}
where $P_0$ is the average pressure
\begin{equation}
\label{E:P0def}
	P_0 = \frac{1}{2} \left( P_L+P_R \right)\,,
\end{equation}
and $\rho$ represents the density matrix for the steady state. The expressions for $J$ and $P$ in the other branch of the steady state solution is rather long and has been omitted. A plot of $J$ and $P$ as a function of $\delta p$ for both branches appears in figure \ref{F:3dsol}. The reader is referred to \cite{Chang:2013gba} for details.

The late time steady-state prediction described in \eqref{E:redbranch} has been verified explicitly for systems whose time evolution is determined by hydrodynamics \cite{Joetalk,Chang:2013gba,Bhaseen:2013ypa}. While the hydrodynamic description is probably a good one at small relative pressure differences, $\delta p \ll 1$, it is unlikely to be valid when the relative pressure difference of the heat baths is large and there are large pressure gradients when one moves away from the steady state towards null infinity. In this work we compare the conjecture of \cite{Joetalk,Chang:2013gba,Bhaseen:2013ypa} to steady states in three dimensional theories which may be described using the AdS/CFT correspondence, up to $\delta p =0.5$, and find excellent agreement.

\section{Constructing black brane steady states}
Simply stated, the AdS/CFT correspondence \cite{Maldacena:1997re,Gubser:1998bc,Witten:1998qj}, or more generally gauge-gravity duality (see, e.g., \cite{McGreevy:2009xe}),
relates a classical $d+1$-dimensional gravitational theory to a planar, $d$-dimensional conformal field theory (CFT) at strong coupling. Of particular importance to the problem at hand is that solutions to classical gravity which follow from the action
\begin{equation}
\label{E:action}
	S = \frac{1}{2\kappa^2}\int \sqrt{g} \left( R + \frac{6}{L_{AdS}^2} \right)d^{4}x
\end{equation}
and asymptote to conformally flat space at spatial infinity hold information about a dual conformal field theory in $3$ space-time dimensions. 

For instance, the black brane solution to the equations of motion which follow from \eqref{E:action} after setting $L_{AdS}=1$ is given by
\begin{equation}
\label{E:blackhole}
	ds^2 =2 dt dr  -r^2 \left(1-\left(\frac{4 \pi T }{3 r} \right)^3 \right) dt^2 + r^2 \left( dx_{\bot}^2 + dz^2 \right)
\end{equation}
with $T$ the Hawking temperature of the black brane.
The solution \eqref{E:blackhole} is dual to a thermal state given by a density matrix $\rho = e^{-\beta H}$ of the conformal field theory, with $H$ its Hamiltonian and $\beta=T^{-1}$ its inverse temperature. The expectation value of the energy momentum tensor of the CFT in this thermal state is given by 
\begin{equation}
	\hbox{Tr} \left(\rho T^{\mu\nu} \right)= \begin{pmatrix} 2P(T) &  0 & 0 \\ 0 & P(T) & 0  \\ 0 & 0 & P(T)  \end{pmatrix}
\end{equation}
with
\begin{equation}
	P(T) = p_0 \left(\frac{4 \pi T}{3} \right)^3 
\end{equation}
where the value of the dimensionless parameter $p_0$ depends on the particular details of the dual theory under consideration. In the (planar and strongly coupled) ABJM theory \cite{Aharony:2008ug}, 
\begin{equation}
	p_0 = \frac{2 N^2}{9\sqrt{2 \lambda}}
\end{equation}
with $N$ the rank of the gauge group and $\lambda = N/k$ with $k$ the Chern-Simons level.

In order to generate a steady state in the dual CFT of the type to which the conjecture of \cite{Joetalk,Chang:2013gba,Bhaseen:2013ypa} applies, we  consider black branes whose dual energy momentum tensor will satisfy
\begin{subequations}
\label{E:Tmnasymptotics}
\begin{equation}
	\lim_{\substack{ z\to\infty \\ t,x_{\bot} \hbox{\tiny finite} } } \hbox{Tr} \left(\rho T^{\mu\nu}(t,x_{\bot}, z) \right)  =  \begin{pmatrix} 2 P(T_R) & 0 & 0 \\ 0 & P(T_R) & 0 \\ 0 & 0 & P(T_R) \end{pmatrix}
\end{equation}
and
\begin{equation}
	\lim_{\substack{ z\to -\infty \\ t,x_{\bot} \hbox{\tiny finite} } } \hbox{Tr} \left(\rho T^{\mu\nu}(t,x_{\bot}, z) \right) =  \begin{pmatrix} 2 P(T_L) & 0 & 0 \\ 0 & P(T_L) & 0 \\ 0 & 0 & P(T_L) \end{pmatrix}  
\end{equation}
\end{subequations}
with $T_R$ and $T_L$ the temperatures of the right and left heat baths respectively and $\rho$ the density matrix of the system. Geometrically, the conditions \eqref{E:Tmnasymptotics} correspond to a metric which asymptotes to \eqref{E:blackhole} as $z$ approaches $\pm\infty$ with corresponding temperature $T_L$ and $T_R$. An initial configuration for the metric which interpolates between the two asymptotic black brane solutions will evolve in time and a time independent configuration is expected to be generated at late times and finite $z$. Once  such a steady state is made available then the energy momentum tensor associated with the late time geometry of such a black brane can be compared to the prediction of \cite{Joetalk,Chang:2013gba,Bhaseen:2013ypa}. This is the strategy we will follow in the remainder of this work. Other examples of out of equilibrium stationary black hole configurations can be found in \cite{Fischetti:2012ps, Fischetti:2012vt, Figueras:2012rb, Emparan:2013fha}.

Consider the metric ansatz
\begin{equation}
\label{E:ansatz}
	ds^2 = 2 dt\left(dr - A(t,z,r) dt - F(t,z,r) dz \right) + \Sigma^2(t,z,r) \left(e^{B(t,z,r)} dx_{\bot}^2 + e^{-B(t,z,r)} dz^2 \right)\,.
\end{equation}
As pointed out in \cite{Chesler:2013lia} this ansatz is invariant under $r \to r+\xi(t,z)$. We will fix this residual diffeomorphism symmetry shortly. 
The equations of motion which follow from \eqref{E:action} take the form
\begin{subequations}
\label{E:allEOM}
\begin{align}
\label{E:SigmaEOM}
	4 \partial_r^2 \Sigma + \Sigma \left(\partial_r B\right)^2 &= 0 \\
\label{E:FEOM}
	-\Sigma^2 \partial_r^2 F - \Sigma^2 \partial_r B\, \partial_r F + C_F[B,\,\Sigma] F & = S_F[\Sigma,\,B] \\
\label{E:SdEOM}
	4 \Sigma^3 \partial_r \dot{\Sigma} + 4 \Sigma^2 \partial_r \Sigma \dot{\Sigma} &= S_{\dot{\Sigma}}[\Sigma,\,B,\,F] \\
\label{E:BdEOM}
	4 \Sigma^4 \partial_r \dot{B} + 4 \Sigma^3 \partial_r \Sigma \dot{B} & = S_{\dot{B}}[\Sigma,\,B,\,F,\,\dot{\Sigma}]  \\
\label{E:AEOM}
	2 \Sigma^4 \partial_r^2 A & = S_A[\Sigma,\,B,\,F,\,\dot{\Sigma},\,\dot{B}]
\end{align}	
\end{subequations}
and
\begin{subequations}
\label{E:extra}
\begin{align}
	\ddot{\Sigma} & = Q_{\ddot{\Sigma}}[\Sigma,\,B,\,F,\,A] \\
	\partial_z\dot{F} & = Q_{\partial_z \dot{F}}[\Sigma,\,B,\,F,\,A]\,,
\end{align}
\end{subequations}
with the following definitions. Dotted variables are given by
\begin{equation}
\label{E:dotX}
	\dot{X} = \partial_t X + A \partial_r X
	\qquad
	\ddot{X} = \partial_t \dot{X} + A \partial_r \dot{X}\,.
\end{equation}
The expressions $C_F$ and $S_X$ in \eqref{E:allEOM} depend only on spatial derivatives of their arguments, and the expressions $Q_X$ in \eqref{E:extra} depend on spatial and time derivative of their arguments. 

The asymptotic AdS boundary, located at $r\to\infty$, is given by the line element
\begin{equation}
\label{E:asymptoticmetric}
	ds^2 =\left( 2 dt dr  + r^2  \left(- dt^2 +  dx_{\bot}^2 + dz^2 \right) \right) \left(1 + \mathcal{O}\left( r^{-1} \right) \right)\,.
\end{equation}
The energy momentum tensor of the dual theory can be read off of the large $r$ asymptotics of \eqref{E:ansatz} as follows. Given \eqref{E:asymptoticmetric} we find that, under the equations of motion, the large $r$ expansion of the metric components $A$, $F$, $\Sigma$ and $B$ are given by
\begin{align}
\begin{split}
\label{E:asymptotics}
	A &  = \frac{1}{2} \left(r+\xi(t,z)\right)^2 - \partial_t \xi(t,z) + \frac{a_1(t,z)}{r+\xi(t,z)} + \mathcal{O}\left( r^{-2} \right) \\ 
	F &  = -\partial_z \xi(t,z) + \frac{f_1(t,z)}{r+\xi(t,z)} + \frac{3 \partial_z b_3(t,z)}{4(r+\xi(t,z))^2} + \mathcal{O}\left( r^{-3} \right)  \\
	\Sigma & = r+\xi(t,z) - \frac{ 3 b_3(t,z)}{40(r+\xi(t,z))^5} + \mathcal{O}\left( r^{-6} \right)  \\
	B & = \frac{b_3(t,z)}{(r+\xi(t,z))^3} + \mathcal{O}\left( r^{-4} \right) \,.
\end{split}
\end{align}
Here $a_1(t,z)$, $f_1(t,z)$ and $b_3(t,z)$ must satisfy the constraints
\begin{equation}
\label{E:conservation1}
	\partial_t a_1 = \frac{3}{4} \partial_z f_1
	\qquad
	\partial_t f_1 = \frac{2}{3} \partial_z a_1 + \partial_z b_3
\end{equation}
(which are a result of \eqref{E:extra}), but are otherwise undetermined. Once \eqref{E:conservation1} are satisfied then \eqref{E:allEOM} ensure that \eqref{E:extra} will hold. 
Following \cite{Balasubramanian:1999re,deHaro:2000xn} the energy momentum tensor of the dual theory is given by
\begin{equation}
\label{E:boundaryTmn}
	\hbox{Tr} \left(\rho T^{\mu\nu} \right)=p_0  \begin{pmatrix} -2 a_1 & \frac{3}{2} f_1 & 0 \\ \frac{3}{2} f_1 & -a_1 - \frac{3}{2} b_3 &  0 \\ 0 & 0 & - a_1 +\frac{3}{2} b_3 \end{pmatrix}\,,
\end{equation}
where $\rho$ is the density matrix associated with the dynamical black brane solution.
In order to generate the steady state \eqref{E:Tmnasymptotics} we will impose
\begin{equation}
	p_0 a_1(t,z\to\infty) = -P(T_R) \qquad
	p_0 a_1(t,z\to-\infty) = -P(T_L)
\end{equation}
and also the initial conditions
\begin{equation}
	f_1(t=0,z\to\pm\infty) = 0 \qquad
	b_3(t=0,z\to\pm\infty) = 0\,.
\end{equation}
which ensure, by causality, that $f_1$ and $b_3$ vanish at large $|z|$ for any $t$.

Our strategy for solving \eqref{E:allEOM} is identical to that presented in \cite{Chesler:2013lia}. At $t=0$ we set 
\begin{equation}
\label{E:inic}
	B=0 \qquad
	f_1=0 \qquad
	a_1 = -A_0 \left(1 - \alpha \tanh \left(\beta \tanh\left(\frac{z}{\lambda}\right) \right) \right)\,.
\end{equation}
Using \eqref{E:inic} and fixing the residual diffeomorphism symmetry parameter $\xi(t)$ we can now determine $\Sigma$, $F$, $\dot{\Sigma}$, $\dot{B}$ and $A$ at $t=0$ by solving \eqref{E:allEOM} in sequence together with the asymptotic conditions \eqref{E:asymptotics}. From this data, one can determine $B$, $a_1$ and $f_1$ at $t=\Delta t$ using \eqref{E:dotX} and \eqref{E:conservation1}. This procedure can now be repeated to obtain the fields at $t=2\Delta t$ and so on, ad infinitum.

The numerical scheme which was implemented in order to evolve the metric forward in time was based on pseudo-spectral methods to solve the linear equations \eqref{E:allEOM} and 3rd order Adams-Bashforth to evolve forward in time. We used a grid of 23 points along the radial direction, $r$, and 100 points along the $z$ direction which we parametrically compactified using 
\begin{equation}
\label{E:ztozeta}
	\zeta = \tanh (z/(\lambda L))
\end{equation}
for an optimal value of $L$. Rescaling all the boundary coordinates by a factor of $\lambda$ is a symmetry of the equations of motion due to conformal invariance of the boundary theory. In practice we have implemented a code which uses $\lambda=1$. In solving the equations of motion we found it convenient to subtract from $A$, $\dot{B}$, $\Sigma$, $\dot{\Sigma}$ and $F$ the solution to \eqref{E:allEOM} at $t=0$ with the initial conditions \eqref{E:inic} which can be obtained analytically, viz.,
\begin{align}
\notag
		\Sigma &= r+\xi &
		F & = -\partial_z \xi &
		A & = \frac{1}{2} (r+\xi)^2 - \partial_t \xi + \frac{a_1}{r+\xi} \\
		\dot{\Sigma} & = \frac{1}{2}(r+\xi)^2 + \frac{a_1}{r+\xi} &
		\dot{B} & = 0\,. &&
\end{align}
In practice the value of $\xi$ was chosen so that an apparent horizon will be located at $r=1$ which is the endpoint of our grid. We refer the reader to \cite{Chesler:2013lia} for details of such a scheme. 

\begin{figure}[hbt]
\centering{
\includegraphics[width=0.495\textwidth]{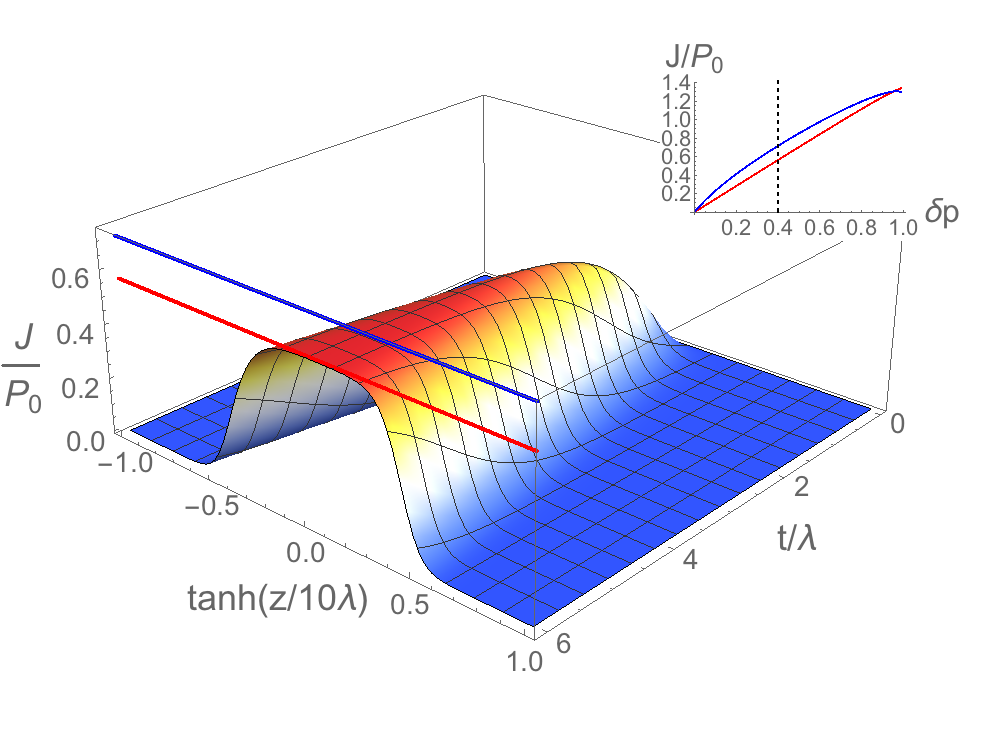}\hfill
\includegraphics[width=0.495\textwidth]{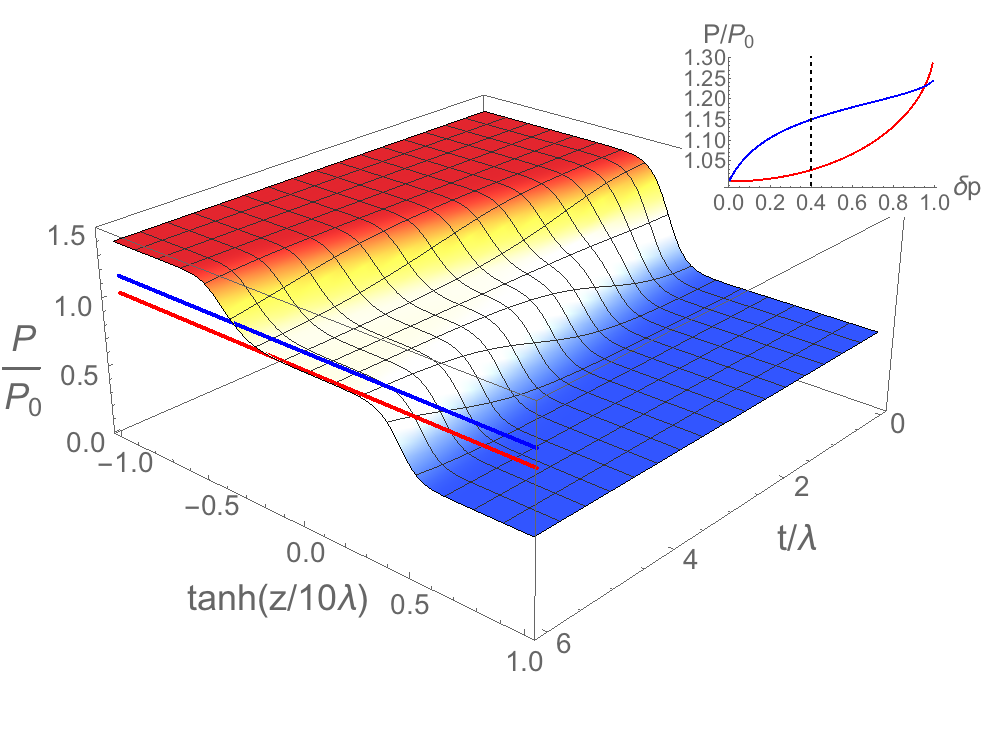}
\caption{ \label{F:Tmn's} A typical plot of the energy flux (left) and pressure (right) of a theory dual to the black hole described by \eqref{E:ansatz}. These plots were generated by solving \eqref{E:allEOM} with initial conditions given in \eqref{E:inic} with $\lambda=1$, $A_0=100$, $\alpha\sim 0.8657$ and $\beta=1/2$ (corresponding to $\delta p = 0.4$). At late times the steady state value of the energy flux and pressure asymptote to the red branch to  within $0.2\%$. }
}
\end{figure}

\begin{figure}[hbt]
\centering{
\includegraphics[width=0.495\textwidth]{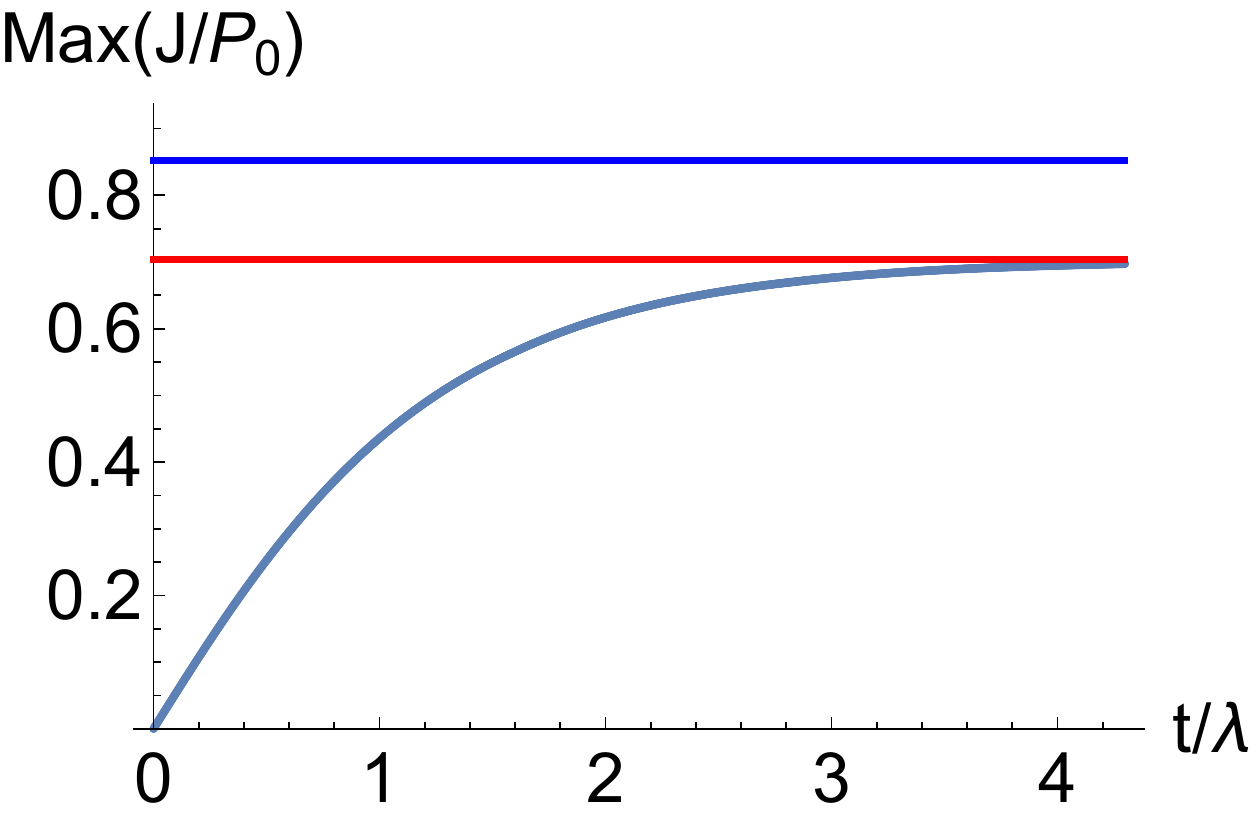}\hfill
\includegraphics[width=0.495\textwidth]{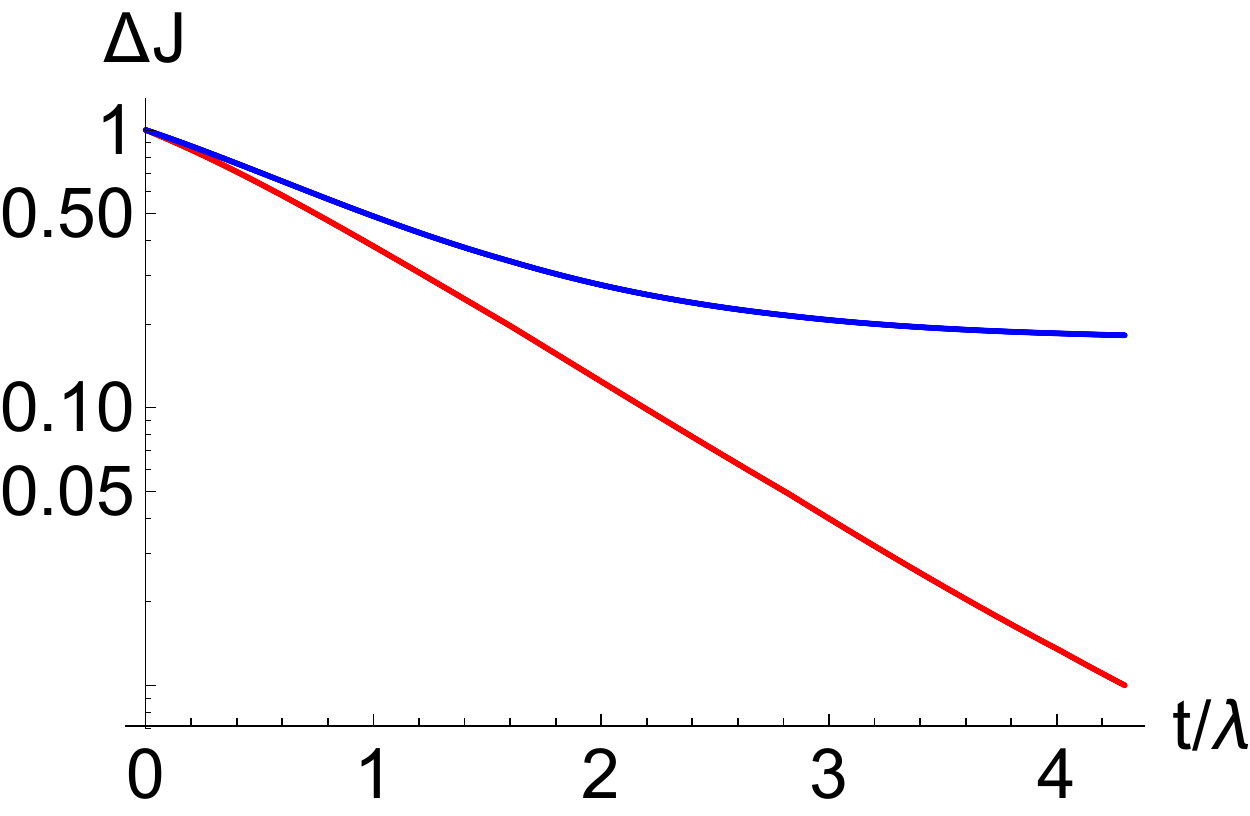}
\caption{ \label{F:approach} Plots describing the approach to the conjectured steady state configuration. These plots were generated by solving \eqref{E:allEOM} with initial conditions given in \eqref{E:inic} with $\lambda=1$, $A_0=100$, $\alpha\sim 1.082$ and $\beta=1/2$ (corresponding to $\delta p = 0.5$). The $z$ coordinate was parametrically compactified as in \eqref{E:ztozeta} with $L=5$. In the left panel we have plotted, for every time $t$, the maximal (in $z$) value for the energy flux $J$ together with the conjectured steady state value of $J$ (red and blue). In the right panel we have plotted the relative difference between the maximal (in $z$) value for the energy flux $J$ and the predicted values of the thermodynamic (red) and other (blue) branch. }
}
\end{figure}

The final expression for the energy momentum tensor of the dual CFT has been computed using \eqref{E:boundaryTmn}. We have carried out the computation for several values of $0<\delta p<0.7$. A typical result is plotted in figure \ref{F:Tmn's}. 
For $\delta p \lesssim 0.5$ we have managed to evolve our numerics for sufficiently long so that the energy flux approached its equilibrium value to an accuracy of less than 1\%.
See figure \ref{F:approach}. 
At very large relative pressure difference the black brane configuration at $z\to\infty$ has a very small temperature relative to the one on at $z\to-\infty$ and our numerical scheme seems to break down. At best this breakdown is due to the parametric compactification we have used to map the $z$ coordinate to a finite interval. In order to avoid difficulties associated with such compactifications \cite{boyd2013chebyshev}, it may be possible to repose the steady state problem using periodic boundary conditions in the $z$ direction. However, if the numerical instability associated with large relative pressure differences is due to the formation of caustics, then one would need a more robust algorithm than that of \cite{Chesler:2013lia} and summarized here to obtain a handle on the problem. 

When $\delta p^2 \sim 0.9$ the steady state pressure and energy density of the other (blue) branch becomes lower than that of the thermodynamic (red) branch. It would be interesting to check if at such large pressure differences the resulting steady state prefers the other branch.

\section*{Acknowledgements}
We thank K. Balasubramanian, H.-C. Chang, P. Chesler, A. Karch and K. Schalm  for useful discussions and comments. AY and IA are supported by the ISF under grant numbers 495/11, 630/14 and 1981/14, by the BSF under grant number 2014350, by the European commission FP7, under IRG 908049 and by the GIF under grant number 1156/2011.

\begin{appendix}

\end{appendix}

\bibliographystyle{JHEP}
\bibliography{BHSS}

\end{document}